\documentclass[12pt]{iopart}
\DeclareUnicodeCharacter{2212}{\textminus}
\DeclareUnicodeCharacter{02B9}{'}

\expandafter\let\csname equation*\endcsname\relax

\expandafter\let\csname endequation*\endcsname\relax

\usepackage{iopams}  
\usepackage{graphicx, subcaption}
\usepackage{siunitx}
\usepackage{physics}
\usepackage{ragged2e}
\justifying

\usepackage{tikz}
\usepackage{tikz-3dplot}
\usepackage{miller}
\usepackage{siunitx}
\usepackage{comment}
\usepackage{booktabs,tabularx}
\usepackage{subcaption}
\usepackage{amssymb}
\usepackage{amsmath}
\usepackage{xcolor}
\usepackage{amssymb}

\begin{document}

\newpage
\title[Influence of Interface Energy Anisotropy on the Solid-state Instability]{Influence of Interface Energy Anisotropy on the Solid-state Instability in Ni-based Superalloy: \\ A Multiscale Study}

\author{Sourav Ghosh$^{a,b,c}$, Christian Brandl$^{b}$, Rajdip Mukherjee$^{a}$}

\address{$^{a}$ Department of Materials Science and Engineering, Indian Institute of Technology Kanpur, Kanpur- 208016, India}
\address{$^b$Department of Mechanical Engineering, The University of  Melbourne, Parkville 3010, Australia}

\address{$^c$School for Engineering of Matter, Transport and Energy, Arizona State University, Tempe 85281, USA}

\ead{rajdipm@iitk.ac.in}
\vspace{10pt}


\begin{abstract}

The microstructural stability of nickel-based superalloys critically depends on the morphology and evolution of $\gamma^\prime$-precipitates, which is governed by elastic and interfacial anisotropies at the atomic scale. 
Here, we present a novel quantitative multiscale framework that, for the first time, directly incorporates atomistically computed interface energy anisotropy into mesoscale phase-field simulations to elucidate morphological selection and instability in the Ni–Al system.

We employ density functional theory (DFT) to accurately predict the orientation-dependent $\gamma/\gamma'$ interface energies for key crystallographic planes. A rigorous analytic mapping is then developed to systematically reduce the three-dimensional (3D) interface anisotropy landscape to the two-dimensional (2D) simulation plane. This enables quantitative transfer of DFT-informed anisotropy parameters into a continuum phase-field model that also accounts for elastic inhomogeneity and eigenstrain.

Our simulations demonstrate that the explicit inclusion of DFT-based interface energy anisotropy fundamentally alters precipitate morphological evolution, robustly suppressing instability and faceting phenomena otherwise promoted by supersaturation and elastic effects. The framework bridges atomic- to mesoscale modeling, enabling predictive control of precipitate shapes and providing new insights into the interplay of elastic and interfacial contributions in Ni-based superalloys. This approach paves the way for quantitative microstructural design in advanced high-temperature alloys via first-principles guided multiscale simulation.

\end{abstract}



\section{Introduction}

Morphological instability of coherent $\gamma/\gamma'$ interfaces in Ni-based superalloys has attracted considerable attention because it governs precipitate shape evolution, directional coarsening, and the formation of dendritic or highly faceted microstructures during thermal exposure~\cite{Yeom1993TheCoarsening, Yoo1995TheSuperalloy, Ghosh2024EffectStudy, Khachaturyan1995ElasticSolids}. Previous theoretical and experimental studies have established that supersaturation, coherency misfit, and elastic anisotropy can destabilize initially equiaxed precipitates, promoting directional growth and interface grooving along crystallographically preferred directions~\cite{Voorhees1992OnSolids, Leo2001OnInstability, Thompson1994ThePrecipitate}. Linear stability analyses have successfully identified the conditions under which elastic interactions amplify interfacial perturbations and trigger morphological instability. However, these studies predominantly attribute instability to elastic effects and provide limited insight into the mechanisms that may counteract or suppress such instabilities.

From a thermodynamic perspective, the evolution of a coherent interface is controlled not only by elastic strain energy but also by the orientation dependence of the interface energy, $\alpha(\vec{n})$, and the corresponding interface stiffness, $\alpha''(\vec{n})$~\cite{Du_D, Ghosh2025InterfacePerspective}. The quantity $(\alpha+\alpha'')_{\vec{n}_0}$ determines the energetic penalty associated with interface perturbations and therefore governs the stability of an interface against shape fluctuations~\cite{Cahn1958FreeEnergy}. Although interface energy anisotropy is known to influence equilibrium crystal shapes and growth morphologies in numerous material systems, its role in solid-state morphological instability of coherent $\gamma/\gamma'$ interfaces in Ni-based superalloys remains largely unexplored. In particular, it is presently unknown whether realistic interface energy anisotropy can suppress elastic-instability-driven dendritic growth, modify the critical conditions for instability onset, or alter the eventual equilibrium morphology of $\gamma'$ precipitates.

Moreover, the few studies that have incorporated interfacial anisotropy in phase-field simulations have generally relied on phenomenological orientation functions or mathematically convenient anisotropy parameters~\cite{Kaptay2012OnInterfaces, Abinandanan2001AnAnisotropy}. Such descriptions may not capture the quantitative anisotropy of the $\gamma/\gamma'$ interface and therefore cannot provide predictive insights into the competition between elastic and interfacial driving forces~\cite{Philippe2022RegularizedSurfaces}. To date, no study has systematically integrated first-principles-derived $\gamma/\gamma'$ interface energy anisotropy with quantitative phase-field simulations to investigate its influence on the initiation, evolution, and stabilization of solid-state morphological instabilities in Ni-based superalloys.

To address this knowledge gap, we develop a quantitative multiscale framework that explicitly incorporates atomistically determined interface energy anisotropy into a phase-field description of coherent precipitate evolution. By combining density functional theory (DFT) calculations of orientation-dependent $\gamma/\gamma'$ interfacial energies with continuum simulations that include both elastic and interfacial anisotropies, we systematically investigate the evolution of precipitate morphologies in a model Ni--Al system under varying supersaturation and misfit strain. The present work reveals how realistic interface energy anisotropy competes with elastic anisotropy to suppress morphological instability, modify the critical aspect ratio for instability onset, and promote more equiaxed equilibrium precipitate morphologies, thereby providing a predictive framework for the design of microstructurally stable high-temperature Ni-based superalloys.

    \begin{figure}
    \centering
    \includegraphics[width=0.5\textwidth]{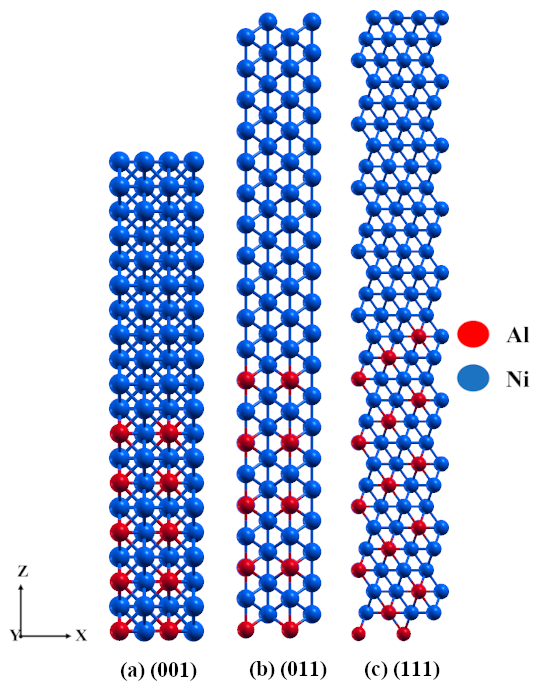}
    \caption{Atomic structure of 3 high symmetric FCC-Ni/ L1\textsubscript{2}-Ni\textsubscript{3}Al  coherent interfaces (a) \hkl(001) (b) \hkl(011) (c) \hkl(111) showing the position of Ni (blue) and Al (red) atoms. The interface plane normal is $\parallel$ to $z$-direction }
    \label{fig:DFT-interface-structure}
\end{figure}

\section{Simulation methodology}

\subsection{First-principles calculations of interface energy}

Ab-initio calculations were performed to predict the interfacial energy of three high-symmetry Ni/Ni$_3$Al interfaces---$(001)$, $(011)$, and $(111)$---using Quantum Espresso~\cite{Scandolo2005}. The spin-polarized generalized gradient approximation (GGA) with the Perdew–Burke–Ernzerhof (PBE) functional~\cite{Perdew1997} was used for the exchange-correlation interaction; the valence configurations were $[Ar]\,3d^8\,4s^2$ and $[Ne]\,3s^2\,3p^1$ for Ni and Al, respectively.

A Monkhorst-Pack $k$-point mesh was adopted for Brillouin zone sampling, adapted to each interface and supercell size; see Table~\ref{Table: k-point-mesh}. 
The cut-off energy was 558~eV. Convergence was accepted when the total energy changed by less than $10^{-5}$~eV/atom and all forces were below $10^{-3}$~eV/\AA{}.

\begin{table}[hbt!]
\caption{K-point mesh used for $\alpha$ $(\mathrm{mJ/m}^{2})$ predictions of three high-symmetry interfaces.}
\centering
\begin{tabular}{l c}
\toprule
Interface & K-point mesh \\
\midrule
$(001)$ & $5 \times 9 \times 1$ \\
$(011)$ & $9 \times 7 \times 1$ \\
$(111)$ & $4 \times 7 \times 2$ \\
\bottomrule
\end{tabular}
\label{Table: k-point-mesh}
\end{table}

For each orientation, two single-crystal supercells with the same number of atomic layers were created in VESTA~\cite{Momma2008VESTA:Analysis} and combined to form a bicrystal supercell. 
The in-plane lattice parameter was set to 3.57~\AA{}, ensuring full coherency at the Ni/Ni$_3$Al interface, as is physically relevant in early-stage precipitation and as commonly assumed in classic studies~\cite{Martins2012EnergyStudy, Jokisaari2017}. 
Coherency was chosen as the baseline because, in experimental Ni-based superalloys, the $\gamma$/$\gamma'$ interfaces remain coherent up to large precipitate sizes before misfit dislocations develop \cite{Ta2014DesignStudy}. 
Allowing the interface to become incoherent (via atomistic relaxation and possible dislocation formation) would capture higher-energy, semi-coherent states, but this would not reflect the coherent interface energy that most directly governs microstructural evolution during the early and intermediate aging stages studied here. 
Hence, the present work focuses on the coherent state, but our framework could be extended to incoherent or partially relaxed interfaces by additional DFT calculations, as suggested.

To estimate $\alpha$, the interfacial energy per unit area, we computed the total energies for both the relaxed bicrystal and the component single-crystal supercells, as follows:
\begin{equation}
    \alpha_N = \frac{E_{\mathrm{Bi-crystal}} - \frac{1}{2}(E_{\mathrm{crystal1}} + E_{\mathrm{crystal2}})}{2A},
    \label{eq:interface_energy}
\end{equation}
where $E_{\mathrm{Bi-crystal}}$ is the total energy of the fully relaxed bicrystal supercell (with both atomic positions and cell dimensions relaxed to equilibrium subject to the coherency constraint), and $E_{\mathrm{crystal1}}$ and $E_{\mathrm{crystal2}}$ are the total energies of the single-crystal supercells with identical atomic composition and lateral dimensions as the corresponding regions in the bicrystal (see Fig.~1). 
The denominator $2A$ arises because the bicrystal contains two identical interfaces of area $A$ each, while $E_{\mathrm{crystal1}}$ and $E_{\mathrm{crystal2}}$ refer to single-crystal slabs occupying half the bicrystal's supercell volume each. 
We note that if $E_{\mathrm{crystal1}}$ and $E_{\mathrm{crystal2}}$ were calculated in supercells with the \emph{same} total volume as the bicrystal, the $1/2$ prefactor would be omitted and the denominator would simply be $A$. 
All conventions were checked for internal consistency by explicit calculation. 

To explore the proposed sensitivity to the coherency constraint, we have also performed (when computationally feasible) additional DFT calculations in which the atomic layers adjacent to the interface were allowed to relax into potentially incoherent configurations. 
For these, all atoms (including interfacial and bulk-like regions) were fully relaxed, and interfacial energy was re-evaluated using equation.~\ref{eq:interface_energy}. 
In this case, equation.~\ref{eq:interface_energy} remains valid, since the total energy changes appropriately with atomic positions; however, the physical meaning of the computed $\alpha_N$ then corresponds to a semi-coherent or incoherent interface. 
.

\begin{table}[hbt!]
\caption{Interface energies $\alpha$ $(\mathrm{mJ/m}^2)$ predictions (coherent interfaces unless otherwise noted).}
\centering
\begin{tabular}{lccc}
\toprule
Method & $(111)$ & $(110)$ & $(100)$ \\
\midrule
Reported & 16\textsuperscript{\cite{Mao2012EffectsCalculations}} & 29\textsuperscript{\cite{Mao2012EffectsCalculations}} & 48.1\textsuperscript{\cite{Martins2012EnergyStudy}} \\
Present work & 12.5 & 28.1 & 46.5 \\
\bottomrule
\end{tabular}
\label{tab:DFT_IE}
\end{table}

\subsection{Dimensional reduction and anisotropy mapping}

We employ our previous atomistically-informed interface energy anisotropy model~\cite{Ghosh2025InterfacePerspective}, validated by large-scale molecular dynamics simulations, which yields the anisotropic interface energy in three dimensions as:
\begin{equation}
    \frac{\Bar{\epsilon}(\vec{n})}{\epsilon_0} = 1 + \epsilon_1 \sum_{i=1}^{3} n_i^4 + \epsilon_2 n_1^2 n_2^2 n_3^2 \left( 1 + \sum_{i=1}^{3} n_i^2 \right), \label{eq:pFm_aniso}
\end{equation}
where $\vec{n} = (n_1, n_2, n_3)$ is the three-dimensional unit interface normal, $\epsilon_0$ is the mean 3D interface energy, and $\epsilon_1$, $\epsilon_2$ are shape anisotropy coefficients. Here, $\bar{\epsilon}(\vec{n})$ explicitly depends on the vectorial orientation of the interface normal, not just a scalar index. For each high-symmetry direction, the normal vector $\vec{n}$ is determined from the Miller indices $(hkl)$, such that $n_1 = h/\sqrt{h^2 + k^2 + l^2}$, $n_2 = k/\sqrt{h^2 + k^2 + l^2}$, and $n_3 = l/\sqrt{h^2 + k^2 + l^2}$.

\begin{figure*}[hbt!]
    \centering
    \includegraphics[width=\textwidth]{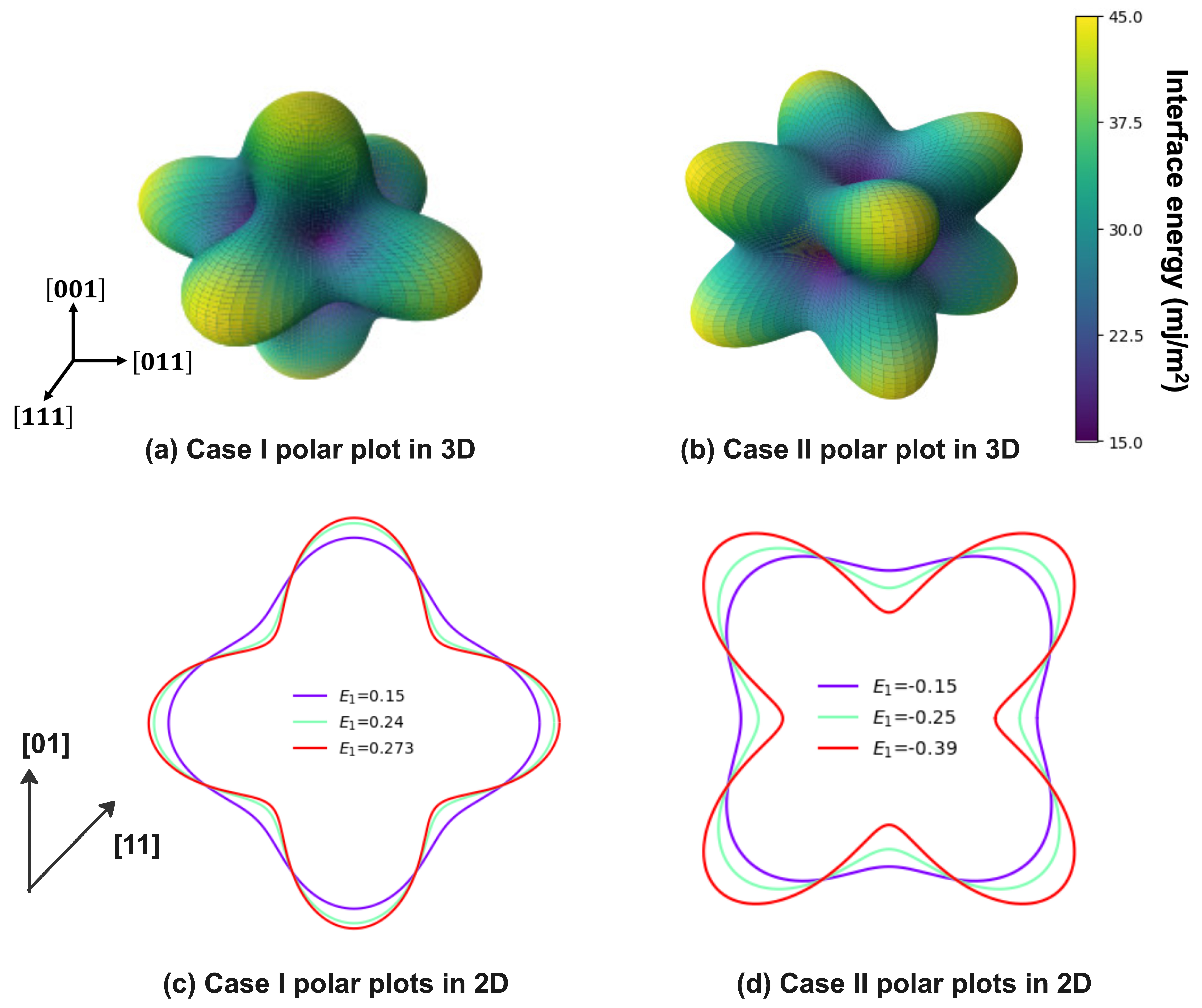}
    \caption{Polar Wulff plots of interface energy anisotropy model predicted from DFT in 3D and 2D.
(a) Case I: 3D polar plot shows Wulff points along the [001] direction and energy cusps along [111].
(b) Case II: 3D polar plot exhibits Wulff points along [111] and cusps along [001], reflecting reversal in anisotropy.
(c) 2D polar plots in the [01]-[11] plane for increasing effective anisotropy ($E_1=0.24$ for 3D prediction).
(d) 2D polar plots for negative values of $E_1$ ($E_1=-0.25$, 3D), illustrating reversal of Wulff shapes versus Case I.
    }
    \label{fig:polar_plot}
\end{figure*}

To connect our DFT-predicted 3D interface energy anisotropy with two-dimensional phase-field simulations, we systematically reduce $\bar{\epsilon}(\vec{n})$ to a function of orientation in the simulation plane. This involves parameterizing the relevant intersection of the Wulff shape with the simulation domain and mapping the 3D anisotropy coefficients $(\epsilon_1, \epsilon_2)$ into an effective 2D anisotropy strength $E_1$ (see Fig.~\ref{fig:polar_plot}  for details). While this mapping incurs some loss of information compared to the full 3D theory, it enables a tractable and physically meaningful assessment of interface energy effects within the available phase-field methodology. We acknowledge that this dimensional reduction introduces an approximation: while the essential features of the anisotropy are retained, certain out-of-plane variations are neglected, and this is discussed as a limitation of the present approach.

In summary, the free energy of the system entering the phase-field model is constructed from DFT-calculated, coherently-relaxed interface energies, using the mapping outlined above to define the orientation-dependent interface energy and its strength for mesoscale simulation.


 . 

Finally, the mean interface energy $\epsilon_0$ in 3D is defined as the area-weighted average of $\alpha_N$ over all orientations,
\begin{equation}
    \epsilon_0 = \frac{1}{4\pi} \int_{\text{all } \vec{n}} \bar{\epsilon}(\vec{n}) \, d\Omega,
\end{equation}
where $d\Omega$ is the solid angle element on the unit sphere. In practice, we numerically evaluate this for the anisotropy model parameters fitted to the DFT data.

Although density functional theory (DFT) calculations provide the three-dimensional orientation dependence of the interfacial energy, all phase-field simulations in this study are performed in two dimensions (2D). This choice is motivated by three key factors: (i) computational feasibility at the large spatial and temporal scales required to observe instability selection and long time shape evolution, (ii) the direct visual interpretability of 2D results—particularly for mapping interface morphologies and characteristic aspect ratios, and (iii) a desire for direct and meaningful comparison with the substantial literature on morphological instabilities, where 2D planar models are frequently benchmarked due to their tractability and ability to expose qualitative and quantitative trends (see, e.g., \cite{Leo1989TheSolution, Mullins1963MorphologicalFlow, Greenwood2009CompetitionDendrites}). 

However, it is critical to note that the 2D anisotropic interface energy in our phase-field simulations is not the outcome of an independent “2D DFT” calculation. Rather, it is rigorously derived from the 3D DFT-based anisotropy model by analytically projecting the three-dimensional dependence onto the 2D simulation plane. 

Concretely, our DFT methodology provides the interface energy as a function of the full interface normal vector $\vec{n} = (n_1, n_2, n_3)$, parameterized according to equation~\ref{eq:pFm_aniso}. In the context of a 2D simulation, we select a physical plane—here, the $(x, y)$ or $(0,0,1)$ plane is used, which corresponds to the plane normal $(0,0,1)$. For any given in-plane interface orientation in simulation, the normal simplifies to $\vec{n} = [\cos\theta, \sin\theta, 0]$, with $\theta$ denoting the angle (measured from the $x$-axis) of the interface normal within the simulation domain. By substituting $n_3 = 0$ into our 3D model, we analytically reduce the orientation dependence to the following 2D expression:
\begin{equation}  
    \frac{\Bar{\epsilon}(\theta)}{\epsilon_0} = 1 + \frac{7}{8}\epsilon_2 + \left(\epsilon_1 + \frac{5}{8}\epsilon_2\right) \cos{4\theta}.
    \label{eq:PFm_aniso_2D}
\end{equation}
Here, all terms that depend on $n_3$ in the 3D model vanish due to the in-plane constraint, and higher-order couplings are accurately remapped into modified anisotropy-strength and offset parameters.

The physical implication of this analytic projection is that our 2D Wulff constructions and interface energy polar plots are \emph{faithful in-plane sections} of the full 3D anisotropy surface: for each $\theta$ in the simulation, the corresponding value of $\Bar{\epsilon}(\theta)$ is precisely the energy cost for an interface oriented at angle $\theta$ within the $(x, y)$ plane, using the full DFT-informed 3D model. The mapping is formalized as:
\begin{align}
    E_1 &= \frac{\epsilon_1 + \frac{5}{8}\epsilon_2}{1 + \frac{7}{8}\epsilon_2}  \label{eq:aniso_strength2D} \\
    E_0 &= \epsilon_0\left(1 + \frac{7}{8}\epsilon_2\right).  \label{eq:aniso_param1_2D}
\end{align}
where $E_1$ is the effective 2D anisotropy strength and $E_0$ is the mean 2D interface energy, fully derived from the DFT parameters for the chosen crystal plane. The corresponding model parameters used in this study are provided in Table~\ref{Table:Interface_aniso_param}.

 A crucial aspect of this approach is that the choice of the two-dimensional simulation plane determines the symmetry and equilibrium morphologies represented in the phase-field calculations. For the present work, the simulations are performed on the $(001)$ plane, and therefore the resulting precipitate shapes and Wulff constructions correspond to the $(001)$ cross-section of the full three-dimensional anisotropic interface-energy surface. In general, different, crystallographically non-equivalent sections, such as the $(110)$ or $(111)$ planes, would produce different two-dimensional projections of the interface-energy anisotropy and consequently different equilibrium morphologies and facet orientations. Thus, the equilibrium shape obtained in a 2D simulation should be interpreted as the projection of the underlying three-dimensional anisotropic interface-energy landscape onto the chosen crystallographic plane. 
Thus, this mapping allows our 2D phase-field results to be directly and quantitatively related to the DFT-derived 3D interface energetics for any desired plane orientation.

 It is important to emphasize that such a dimensional reduction naturally omits all out-of-plane morphological features, for example, three-dimensional facetting, connectivity, and possible singularities associated with the full 3D interface curvature. 
 The essential anisotropy physics are, however, preserved within the plane, making the 2D model predictive for trends and mechanics dominated by in-plane energetics and for direct comparison with experiment and established literature. 
 For truly three-dimensional effects, a full 3D phase-field model using the complete 3D DFT interface energy map would be necessary, albeit at far higher computational cost and complexity. 
 The analytic projection method presented here thus strikes a practical balance, retaining full atomistic fidelity for the studied plane while enabling comprehensive parametric and temporal studies at simulation sizes unattainable for 3D models.
 The 2D anisotropy model in our phase-field framework is not an arbitrary or qualitative assumption, nor an artifact of mathematical convenience; it results from a traceable, analytic restriction of the 3D DFT-informed interface energy to the physics and geometry of the simulation plane of interest, in this work the $(x, y)$-plane normal to $(0,0,1)$.

 The analytic 2D mapping, using Eqs.~(\ref{eq:PFm_aniso_2D})–(\ref{eq:aniso_param1_2D}), enables a direct and physically consistent transfer of atomistic anisotropy information to the mesoscale model.

\begin{table}[hbt!]
\caption{Interface anisotropy parameters for 2D phase-field model, determined from DFT simulation data via equation.~(\ref{eq:PFm_aniso_2D}).}
\centering
\begin{tabular}{lccc}
\toprule
Case & $\epsilon_1$ & $\epsilon_2$ & $E_1$ \\
\midrule
I  & 0.2832   & $-$0.192  & 0.24 \\
II & $-$0.84765 & 0.7391 & $-$0.39024 \\
\bottomrule
\end{tabular}
\label{Table:Interface_aniso_param}
\end{table}

The corresponding 2D polar Wulff plots, constructed using this mapping, are displayed in figure~\ref{fig:polar_plot}(c)–(d). These plots visualize the equilibrium shape that a precipitate would adopt under the influence of the specific in-plane interface energy anisotropy. Our analysis reveals that when the anisotropy is dominated by the \hkl[11] interface normal (Case I), the Wulff point (corresponding to the equilibrium facet orientation) lies along the \hkl[01] direction, while cusps—indicative of energetically less favourable and potentially unstable orientations—are observed along the \hkl[11] direction.

To further elucidate the role of the preferred interface orientation on morphological selection, we also consider a complementary anisotropy scenario (Case II), in which the energetic preference is deliberately reversed such that the \hkl[01] interface normal possesses the lowest interface energy. This hypothetical case is introduced to systematically examine how a reversal in the anisotropic energy landscape alters the predicted equilibrium morphology and to establish the sensitivity of the phase-field evolution to the orientation dependence of the interfacial energy. Under this inverted anisotropy, the Wulff point shifts to the \hkl[11] direction, while the cusps move to the \hkl[01] direction. The reversal of the Wulff and cusp orientations directly reflects the interchange in energetic preference between the two interface normals and demonstrates the predictive capability of the atomistically informed anisotropy framework in capturing orientation-dependent morphological selection.
These behaviors directly reflect the reversal in energetic preference and underline the predictive capability of our atomistically-informed anisotropy model for driving morphological selection in phase-field simulations.

A multiscale modeling strategy is thus realized, wherein the DFT-derived interface anisotropy is quantitatively incorporated into the phase-field simulation via the orientation-dependent gradient energy coefficient. The total free energy functional governing the system evolution takes the form
\begin{equation}
    F = F_{\mathrm{chemical}} + F_{\mathrm{el}},
    \label{Eq:Total_free_ene}
\end{equation}
where $F_{\mathrm{chemical}} = \int_V \left[ f(c(\vec{r})) + \frac{1}{2} \Bar{\epsilon}(\theta) |\nabla c|^2 \right] dV$ comprises a Helmholtz-type chemical free energy density $f(c)$ (chosen here as $A_1 c^2 (1-c)^2$ for computational efficiency), with $\Bar{\epsilon}(\theta)$ conveying the local interface energy density as a function of orientation angle $\theta$. The elastic contribution, $F_{\mathrm{el}}$, is rigorously formulated as detailed above.

The orientation angle at each point is computed according to:
\begin{equation}
    \theta = \tan^{-1}\left( \frac{c_2}{c_1} \right),
    \label{eq:orientation_calc}
\end{equation}
where $c_1$ and $c_2$ represent the spatial derivatives of the conserved composition field in the $x$ and $y$ directions, respectively. The interface energy anisotropy enters as
\begin{equation}
    \Bar{\epsilon}(\theta) = E_0(1 + E_1 \cos 4\theta),
    \label{eq:pf_aniso}
\end{equation}
with $E_0$ and $E_1$ as defined above.

To ensure thermodynamic consistency and correct facet selection, we check the convexity of $1/\Bar{\epsilon}(\theta)$ and identify potential missing orientations—those for which $E_1$ exceeds the threshold for polar plot non-convexity ($E_1 > 1/15$). The critical (corner) angles $\theta_c$ associated with facet formation or forbidden orientations are identified by solving
\begin{equation}
    \Bar{\epsilon}(\theta_c) \sin\theta_c + \Bar{\epsilon}(\theta_c) \cos\theta_c = 0.
    \label{eq:anisotropy_egg_condition}
\end{equation}
When necessary, the gradient energy coefficient is regularized according to Eggleston’s method~\cite{Eggleston2001AEnergy}:
\begin{equation}
    \Bar{\epsilon}(\theta) = \Bar{\epsilon}(\theta_c) \frac{\cos\theta}{\cos\theta_c}
\end{equation}
for $\theta < \theta_c$ and in equivalent formulations for other quadrants, ensuring local equilibrium and smooth shape evolution.

The simulations employ periodic boundary conditions in both spatial directions. This ensures the absence of artificial boundary effects, reflecting the microstructure as an infinite periodic repetition of the simulation domain. This choice is both standard for instability studies and essential for capturing the intrinsic evolution and selection of morphologies without interference from wall or finite-size effects.

The phase-field model further incorporates realistic elastic energy contributions accounting for inhomogeneous elasticity and misfit, as described earlier, with simulation details summarized in Table~\ref{Table: Numerical_param}. The evolution of the composition field is solved using a semi-implicit Fourier spectral method for computational stability and efficiency.

\begin{table}[hbt!]
\caption{Summary of the numerical details used in the phase-field simulations.}
\centering
\begin{tabular}{l c}
\toprule
Parameter & Value / Unit \\
\midrule
System size  & $2048 \times 2048$ grid points \\
Grid spacing  & $0.01$ (dimensionless) \\
Time step  & $0.125$ (dimensionless) \\
Atomic mobility $M$ & $1.0$ (normalized) \\
Shear modulus in $\gamma$-phase, $G^{\gamma}$ & $56.32$ GPa \\
Shear modulus in $\gamma'$-phase, $G^{\gamma'}$ & $61.95$ GPa \\
Zener anisotropy in $\gamma$-phase, $A_Z^{\gamma}$ & $3.59$ \\
Zener anisotropy in $\gamma'$-phase, $A_Z^{\gamma'}$ & $2.69$ \\
Misfit strain, $\epsilon_c$ & $0.0085$ \cite{Ghosh2024EffectStudy}\\
\bottomrule
\end{tabular}
\label{Table: Numerical_param}
\end{table}

\begin{figure*} [hbt!]
    \centering
         \includegraphics[width=0.9\textwidth]{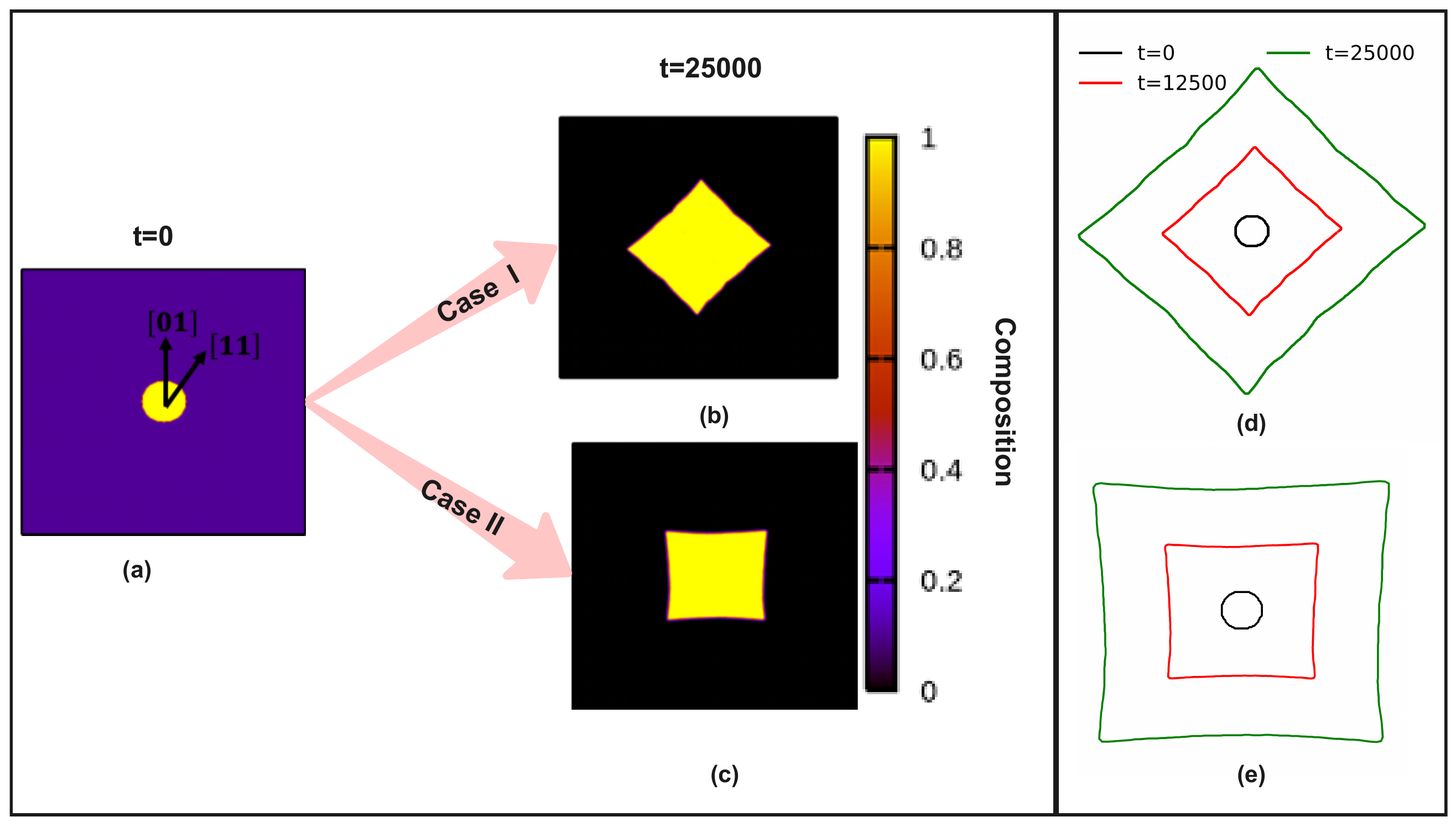}
    \caption{Temporal evolution of a circular $\gamma'$ particle ($a$) attaining the near-equilibrium shape in Case~I ($b$) and Case~II ($c$), respectively. The corresponding composition contours at $t = 0$ and $t = 25{,}000$ time steps are shown for Case~I ($d$) and Case~II ($e$).}
    \label{fig:equillibrium_shape_interface_aniso1}
    \end{figure*}
\section{Results $\&$ Discussions}
 To characterize the effect of Case~I and Case~II interface anisotropy on precipitate morphology,  the temporal evolution of a circular $\gamma^\prime$-precipitate growing from a $3.5~\%$ supersaturated solid solution has been investigated in fig.~\ref{fig:equillibrium_shape_interface_aniso1} at high interface energy anisotropy for present work (see table.~\ref{Table:Interface_aniso_param}). \textcolor{black}{Due to the absence of volume preservation techniques, the final morphologies represent kinetically stabilized shapes, often referred to as "kinetic Wulff shapes" rather than true equilibrium shapes, as discussed in prior works \cite{Eggleston2001AEnergy, Qin2009Phase-fieldMetals, Roy2017InterfacialPrecipitates}.}

Analysis of fig.~\ref{fig:equillibrium_shape_interface_aniso1}$(a)-(b)$  reveals that the $\gamma^\prime$-precipitate stabilizes with faceting in the \hkl[01] direction, as indicated by the Wulff plot in fig.~\ref{fig:polar_plot}(c). This morphology arises due to the lowest interface energy being aligned with the [11] direction. Similarly, fig.~\ref{fig:equillibrium_shape_interface_aniso1}$(c)-(d)$ shows faceting in the [11] direction, as predicted by the Wulff plot in fig. fig.~\ref{fig:polar_plot}(d), where the lowest interface energy is in the [01] direction. These findings confirm that faceting is dominated by the direction of the lowest interface energy (the "softest" direction), consistent with observations from prior studies \cite{Eggleston2001AEnergy, Qin2009Phase-fieldMetals, Roy2017InterfacialPrecipitates}. This investigation highlights the influence of quantitative interface anisotropy on stabilized precipitate morphologies and provides a framework for understanding how these effects interact with elastic anisotropy. For the remainder of this article's phase-field study, we will focus on Case~I interface anisotropy, which aligns with atomistic predictions for Ni-based superalloys. This approach is chosen to demonstrate a quantitative effect of interface anisotropy, in the presence of elastic anisotropy, on precipitate morphology using our multiscale methodology.

\begin{figure*} [hbt!]
    \centering
         \includegraphics[width=0.8\textwidth]{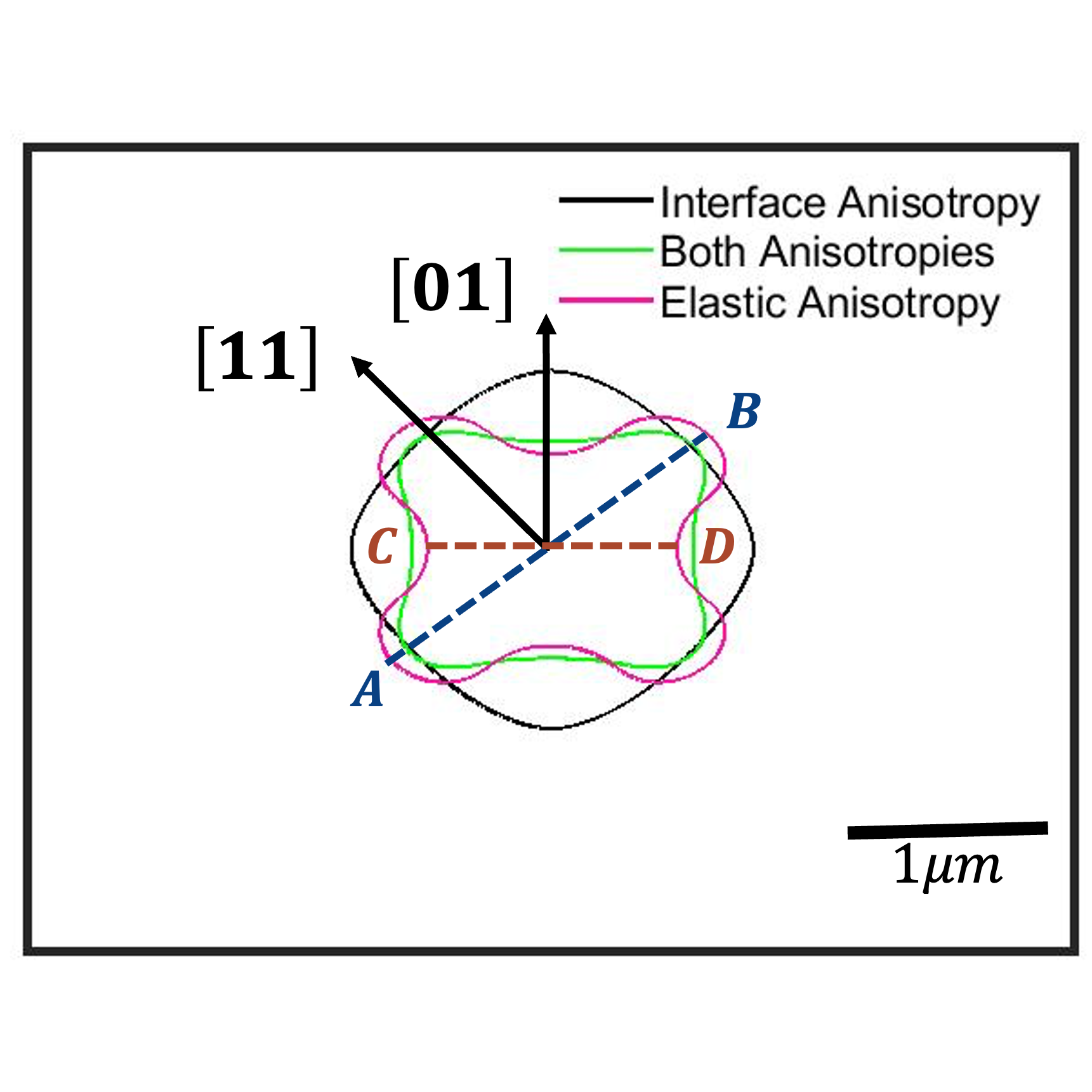}
    \caption{Contour plot (at $\phi=0.5$) of (a) a $\gamma^\prime$-particle at growth regime describing the effect of interface anisotropy (black), both anisotropies (elastic and interface anisotropy) (green) and elastic anisotropy (purple). $AB$ and $CD$ define the major axis (blue)  and the minor axis  (brown), respectively}
    \label{fig:equillibrium_shape_interface_aniso}
    \end{figure*}

    The individual effect of interface anisotropy, elastic anisotropy, and both anisotropies on precipitate shape evolution has been demonstrated in fig.~\ref{fig:equillibrium_shape_interface_aniso}. The following investigation reveals that faceting is observed in [01]-direction in $\gamma\prime$-precipitate (black line) due to the effect of interface anisotropy (Case~I). Further investigation reveals that elastic anisotropy drives the precipitate (purple line) faceting along \hkl[11] direction and grooving along the \hkl[01] direction. The following phenomena were observed due to \hkl[01] being the elastically softest direction in elastic anisotropy, as also shown in fig.~\ref{fig:equillibrium_shape_interface_aniso}. But, grooving along \hkl[01] direction disappears, and the corner along the \hkl[11] direction has been rounded due to the combined effect of \hkl[11] interface-dominated interface anisotropy and elastic anisotropy with elastically soft precipitate. 
\\
Therefore, a more detailed inspection into the contour plot of  $\gamma\prime$-precipitate in the case of elastic anisotropy and both anisotropies reveals that \hkl[11] dominated interface anisotropy is trying to produce faceting along \hkl[01] direction and produce a flat surface along [11] direction which suppresses effect elastic anisotropy to produce the groove formation along \hkl[01] direction and faceting along \hkl[11] direction. To quantify this effect, we compare the temporal evolution of precipitate shape as a function of aspect ratio as a ratio of the length of the major axis (corner to corner distance, $AB$) to the minor axis (groove distance, $CD$) as shown in fig.~\ref{fig:equillibrium_shape_interface_aniso}.

\begin{figure} [hbt!]
    \centering
         \includegraphics[width=\textwidth]{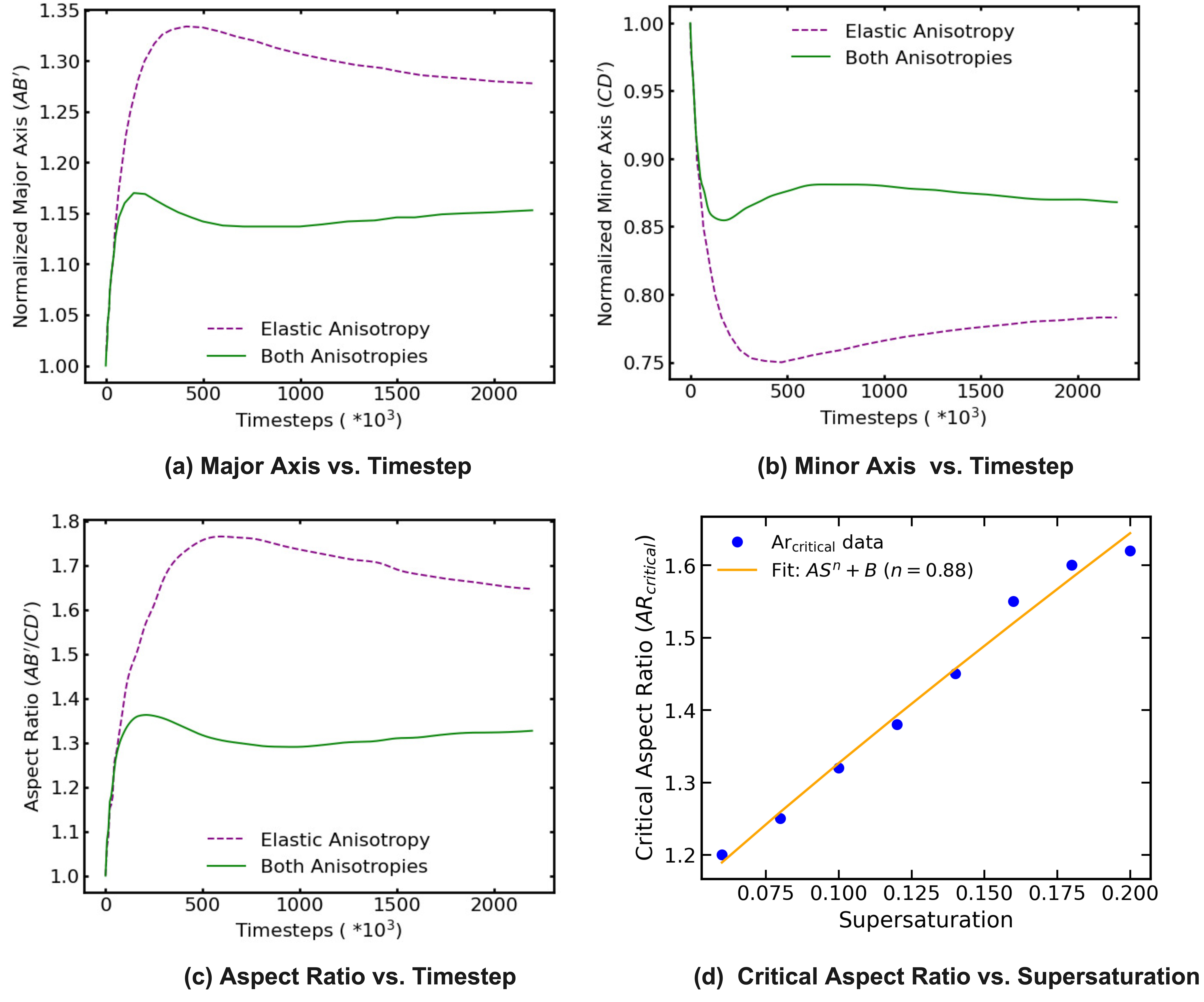}
    \caption{((a) Major axis evolution, (b) minor axis evolution, (c) aspect ratio of $\gamma\prime$-particle during the temporal evolution in case of the effect of  both anisotropies (elastic and interface anisotropy) (green) and elastic anisotropy (purple), and effect of supersaturation on the critical aspect ratio}
    \label{fig:aspect_ratio_two_aniso}
    \end{figure}
    
The temporal evolution of the aspect ratio of $\gamma^\prime$-precipitates is shown in fig.~\ref{fig:aspect_ratio_two_aniso} (c) for the cases of elastic anisotropy alone (purple) and the combined effect of elastic and interface anisotropy (green). In both cases, the aspect ratio initially increases to a maximum value as the precipitate grows, after which growth saturates, signaling either the completion of growth or exhaustion of supersaturation. However, the maximum aspect ratio differs significantly. Under elastic anisotropy, the aspect ratio peaks at $\sim$1.75, while under combined anisotropy, the peak is limited to $\sim$1.35, i.e., almost $23\%$ lower. Furthermore, the time to reach the maximum aspect ratio is reduced by nearly a factor of 2.7 in the combined anisotropy case, and the morphology relaxes to its almost stable aspect ratio (within $1\%$ of the final value) approximately 2.6 times earlier. These quantitative differences clearly demonstrate the strong stabilizing influence of interface anisotropy.

To understand the origin of this behavior, fig.~\ref{fig:aspect_ratio_two_aniso} (a) and (b) show the temporal evolution of the major ($AB^\prime$) and minor ($CD^\prime$) axis, where we have used the normalized lengths $AB^\prime = AB/AB_0$ and $CD^\prime = CD/CD_0$ with respect to their initial values ($AB_0, CD_0$). Under elastic anisotropy, the major axis grows monotonically to a maximum of $AB’ \approx 1.33$, while the minor axis contracts steadily to a minimum of $CD’ \approx 0.75$. This strong elongation along compliant crystallographic directions amplifies the Mullins–Sekerka instability, resulting in a peak aspect ratio of $\sim 1.75$. In contrast, when interface anisotropy is included, the major axis growth is curtailed at $AB’ \approx 1.17$, while the minor axis contracts less severely, reaching only $CD’ \approx 0.87$ before recovering. This combined behavior yields a reduced aspect-ratio peak of $\sim 1.35$, approximately $23\%$ lower than the elastic-only case. For elastic anisotropy, the major axis grows steadily while the minor axis shrinks monotonically, leading to elongation along compliant crystallographic directions and thus amplifying the Mullins–Sekerka instability. The reduction arises directly from the stabilizing contribution of interface anisotropy: the effective interfacial stiffness $\tilde{\gamma}=\gamma+\gamma''$ increases the curvature penalty at the tips (see fig.~\ref{fig:equillibrium_shape_interface_aniso}), suppresses solute/chemical potential focusing, and thus reduces the extent of tip sharpening. Consequently, the effective capillary length is larger, the supersaturation driving force is exhausted earlier, and the precipitate relaxes more quickly to a lower, more equiaxed steady state morphology. We can further observe that during the temporal evolution, there exists a distinct point at which the aspect ratio profile of interface anisotropy begins to deviate from that of the combined (both elastic and interfacial) anisotropy. This point marks the onset of morphological instability and is defined as the critical aspect ratio, ( $\text{AR}_{\text{critical}}$). 
Physically, ($\text{AR}{\text{critical}}$) represents the transition point at which the balance between interfacial and elastic energies breaks down. Below this threshold, the interfacial energy—through its anisotropic stiffness—effectively counteracts elastic-driven elongation, maintaining a stable and nearly equiaxed morphology. Beyond ($\text{AR}{\text{critical}}$), however, the accumulated elastic strain energy exceeds the stabilizing influence of the interface, leading to preferential growth along elastically favorable directions. This marks the onset of shape divergence between the two anisotropy cases and the emergence of morphological instability.
This deviation is intrinsically linked to the effective supersaturation in the system. As supersaturation increases, the chemical potential difference driving precipitate growth becomes stronger, promoting more pronounced elongation and instability. To systematically investigate this, we analyzed the effect of supersaturation on (  $\text{AR}_{\text{critical}}$ ) by tracking the temporal evolution of aspect ratio at several supersaturation values. For each case, (  $\text{AR}_{\text{critical}}$ ) was identified as the aspect ratio where the interface anisotropy profile first diverges from the both-anisotropy profile. Fig.~\ref{fig:aspect_ratio_two_aniso} (d) illustrates $\text{AR}_{\text{critical}}$ as a function of supersaturation, which reveals a clear increasing trend: higher supersaturation produces larger critical aspect ratios, indicating earlier onset and greater magnitude of morphological instability. This trend was quantitatively compared with a power-law model of the form $\text{AR}_{\text{critical}} = AS^n + B$, where $S$ is supersaturation and $ n$ is the growth exponent. Fitting our data yields an exponent $n \simeq 0.88$, suggesting a sublinear yet strong dependence of critical aspect ratio on supersaturation. Such a value lies between classical diffusion-controlled growth ($n=0.5$) and interface-controlled regimes ($n=1$), reflecting the mixed control over precipitate morphology in the presence of both anisotropic effects \cite{Voorhees1992OnSolids}. Physically, this result indicates that while interface anisotropy stabilizes precipitate shape and delays the onset of instability, increasing supersaturation amplifies elastic-driven elongation and tip sharpening, pushing the system more rapidly toward the critical aspect ratio. The fitted exponent underscores the interplay between interface and elastic anisotropies under non-equilibrium growth conditions, providing a quantitative framework for predicting morphological transitions in $\gamma^\prime$-precipitates as a function of supersaturation.  Therefore, elastic anisotropy alone promotes instability and elongation, whereas the inclusion of interface anisotropy strongly suppresses this effect, leading to earlier growth arrest and a lower steady-state aspect ratio.

\section{Conclusions}
In summary, we investigate the morphological instability of the $\gamma/\gamma^\prime$-interface in a model Ni-based superalloy using a quantitative multiscale phase-field approach incorporating elastic and interfacial anisotropies. The following conclusions are drawn:

\begin{enumerate}
    \item The anisotropic elastic energy induces interface faceting and grooving, with facets forming along the ⟨11⟩ direction and grooves along ⟨01⟩, consistent with the directional elastic stiffness \cite{Voorhees1992OnSolids, Leo2001OnInstability}.

    \item Quantitative interface energy anisotropy, as derived from first-principles calculations, effectively suppresses the morphological instability that would otherwise be triggered by supersaturation, lattice misfit, and elastic anisotropy. This leads to more equiaxed and stable precipitate morphologies.

    \item  For a given misfit strain, the interplay between elastic and interface anisotropies dictates not only the shape evolution but also the equilibrium morphology of $\gamma/\gamma^\prime$ precipitates.

    \item The onset and magnitude of morphological instability are set by a critical aspect ratio, $\text{AR}_{\text{critical}}$ , where interfacial anisotropy can no longer offset elastic-driven elongation; adding interface anisotropy lowers the peak aspect ratio and yields a more equiaxed steady state. $\text{AR}_{\text{critical}}$  increases sublinearly with supersaturation following $\text{AR}_{\text{critical}} = AS^{0.88} + B$, indicating earlier and stronger instability at higher supersaturation

    \item  This multiscale atomistic-to-mesoscale framework delivers a physically grounded, predictive methodology for understanding and controlling microstructural instability in Ni-based superalloys, offering new avenues for alloy and processing design for high-temperature performance.

\end{enumerate}

\section*{CRediT authorship contribution statement}
\textbf{Sourav Ghosh}: Writing – review \& editing, Writing – original draft, Visualization, Validation, Software, Methodology, Investigation, Formal analysis, Conceptualisation.
\textbf{Rajdip Mukherjee}: Writing – review \& editing, Supervision, Funding acquisition, Conceptualization.
\textbf{Christian Brandl}: Writing – review \& editing, Supervision, Resources, Software, Investigation, Conceptualization.

\section*{Acknowledgements}
SG has been funded by the Melbourne Research Scholarship (MRS) and the institute fellowship by the Indian Institute of Technology, Kanpur.  We acknowledge the National Supercomputing Mission (NSM) for providing computing resources of "PARAM Sanganak" at IIT Kanpur, which is implemented by C-DAC and supported by the Ministry of Electronics and Information Technology (MeitY) and Department of Science and Technology (DST), Government of India. We also acknowledge the high-performance computing facility (Spartan) and Melbourne Research Cloud (MRC) at the University of Melbourne for providing computing resources.

\section*{Data Availability Statement}
The data that support the findings of this 
study are available from the corresponding 
author upon reasonable request.

\section*{Declaration of competing interest}
The authors declare that they have no known competing financial interests or personal relationships that could have appeared to influence
the work reported in this paper.



\bibliographystyle{unsrt}
\bibliography{references}  

@article{Scandolo2005,
    title = {{First-principles codes for computational crystallography in the quantum-espresso package}},
    year = {2005},
    journal = {Zeitschrift Fur Kristallographie},
    author = {Scandolo, Sandro and Giannozzi, Paolo and Cavazzoni, Carlo and De Gironcoli, Stefano and Pasquarello, Alfredo and Baroni, Stefano},
    number = {5-6},
    pages = {574--579},
    volume = {220}
}

@article{Perdew1997,
    title = {{Generalized gradient approximation made simple (vol 77, pg 3865, 1996)}},
    year = {1997},
    journal = {Physical Review Letters},
    author = {Perdew, John P. and Burke, Kieron and Ernzerhof, Matthias},
    number = {7},
    pages = {1396--1396},
    volume = {78}
}

@article{Jokisaari2017,
    title = {{Predicting the morphologies of {$\gamma$}ʹ precipitates in cobalt-based superalloys}},
    year = {2017},
    journal = {Acta Materialia},
    author = {Jokisaari, A. M. and Naghavi, S. S. and Wolverton, C. and Voorhees, P. W. and Heinonen, O. G.},
    pages = {273--284},
    volume = {141},
    publisher = {Elsevier Ltd}
}

@article{Du_D,
    title = {{Properties and determination of the interface stiffness}},
    year = {2007},
    journal = {Acta Materialia},
    author = {Du, Danxu and Zhang, Hao and Srolovitz, David J.},
    number = {2},
    month = {1},
    pages = {467--471},
    volume = {55}
}

@article{Eggleston2001AEnergy,
    title = {{A phase-field model for highly anisotropic interfacial energy}},
    year = {2001},
    journal = {Physica D},
    author = {Eggleston, J J and Mcfadden, G B and Voorhees, P W},
    pages = {91--103},
    volume = {150}
}

@article{Abinandanan2001AnAnisotropy,
    title = {{An extended Cahn-Hilliard model for interfaces with cubic anisotropy}},
    year = {2001},
    journal = {Philosophical Magazine A: Physics Of Condensed Matter, Structure, Defects And Mechanical Properties},
    author = {Abinandanan, T. A. and Haider, F.},
    number = {10},
    month = {10},
    pages = {2457--2479},
    volume = {81}
}

@article{Greenwood2009CompetitionDendrites,
    title = {{Competition between surface energy and elastic anisotropies in the growth of coherent solid-state dendrites}},
    year = {2009},
    journal = {Acta Materialia},
    author = {Greenwood, Michael and Hoyt, Jeffrey J. and Provatas, Nikolas},
    number = {9},
    month = {5},
    pages = {2613--2623},
    volume = {57}
}

@inproceedings{Ta2014DesignStudy,
    title = {{Design of the precipitation process for Ni-Al alloys with optimal mechanical properties: A phase-field study}},
    year = {2014},
    booktitle = {Metallurgical and Materials Transactions A: Physical Metallurgy and Materials Science},
    author = {Ta, Na and Zhang, Lijun and Du, Yong},
    number = {4},
    pages = {1787--1802},
    volume = {45},
    publisher = {Springer Boston}
}

@article{Ghosh2024EffectStudy,
    title = {{Effect of molybdenum addition on precipitate coarsening kinetics in Inconel 740H: A phase-field study}},
    year = {2024},
    journal = {Intermetallics},
    author = {Ghosh, Sourav and Kumar, Hemanth and Brandl, Christian and Choudhury, Abhik N. and Bhattacharyya, Saswata and Mukherjee, Rajdip},
    month = {12},
    volume = {175},
    publisher = {Elsevier Ltd}
}

@article{Mao2012EffectsCalculations,
    title = {{Effects of temperature and ferromagnetism on the {$\gamma$}-Ni/{$\gamma$}'- Ni3Al interfacial free energy from first principles calculations}},
    year = {2012},
    journal = {Journal Of Materials Science},
    author = {Mao, Zugang and Booth-Morrison, Christopher and Plotnikov, Elizaveta and Seidman, David N.},
    number = {21},
    month = {11},
    pages = {7653--7659},
    volume = {47}
}

@article{Khachaturyan1995ElasticSolids,
    title = {{Elastic strain energy of inhomogeneous solids}},
    year = {1995},
    journal = {Physical Review B},
    author = {Khachaturyan, A G and Semenovskaya, S and Tsakalakos, T},
    number = {15},
    pages = {909--919},
    volume = {52}
}

@article{Martins2012EnergyStudy,
    title = {{Energy of Ni/Ni3Al interface: A temperature-dependent theoretical study}},
    year = {2012},
    journal = {Materials Letters},
    author = {Martins, A. S. and De Campos, M. F.},
    month = {9},
    pages = {100--103},
    volume = {83}
}

@article{Cahn1958FreeEnergy,
    title = {{Free energy of a nonuniform system. I. Interfacial free energy}},
    year = {1958},
    journal = {The Journal Of Chemical Physics},
    author = {Cahn, John W. and Hilliard, John E.},
    number = {2},
    pages = {258--267},
    volume = {28}
}

@article{Ghosh2025InterfacePerspective,
    title = {{Interface energy anisotropy and interface step structure in {$\gamma$}-Ni/{$\gamma$}'−Ni3Al : An atomistic simulations perspective}},
    year = {2025},
    journal = {Physical Review Materials},
    author = {Ghosh, Sourav and Mukherjee, Rajdip and Brandl, Christian},
    number = {10},
    month = {10},
    pages = {103601},
    volume = {9},
    issn = {2475-9953}
}

@article{Roy2017InterfacialPrecipitates,
    title = {{Interfacial free energy anisotropy driven faceting of precipitates}},
    year = {2017},
    journal = {Philosophical Magazine},
    author = {Roy, Arijit and Nani, E S and Lahiri, Arka and Gururajan, M P},
    number = {30},
    month = {10},
    pages = {2705--2735},
    volume = {97}
}

@article{Mullins1963MorphologicalFlow,
    title = {{Morphological stability of a particle growing by diffusion or heat flow}},
    year = {1963},
    journal = {Journal Of Applied Physics},
    author = {Mullins, W. W. and Sekerka, R. F.},
    number = {2},
    pages = {323--329},
    volume = {34}
}

@article{Leo2001OnInstability,
    title = {{On an elastically induced splitting instability}},
    year = {2001},
    journal = {Acta Materialia},
    author = {Leo, P.H. and Lowengrub, J.S. and Nie, Q.},
    number = {14},
    month = {8},
    pages = {2761--2772},
    volume = {49},
    publisher = {Pergamon}
}

@article{Kaptay2012OnInterfaces,
    title = {{On the interfacial energy of coherent interfaces}},
    year = {2012},
    journal = {Acta Materialia},
    author = {Kaptay, G.},
    number = {19},
    month = {11},
    pages = {6804--6813},
    volume = {60}
}

@article{Voorhees1992OnSolids,
    title = {{On the morphological development of second-phase particles in elastically-stressed solids}},
    year = {1992},
    journal = {Acta Metallurgica},
    author = {Voorhees, P W and McFADDEN, G B and Johnson, W C},
    number = {11},
    pages = {2979--2992},
    volume = {40}
}

@article{Qin2009Phase-fieldMetals,
    title = {{Phase-field model study of the effect of interface anisotropy on the crystal morphological evolution of cubic metals}},
    year = {2009},
    journal = {Acta Materialia},
    author = {Qin, R. S. and Bhadeshia, H. K.D.H.},
    number = {7},
    month = {4},
    pages = {2210--2216},
    volume = {57}
}

@article{Philippe2022RegularizedSurfaces,
    title = {{Regularized anisotropic motion-by-curvature in phase-field theory: Interface phase separation of crystal surfaces}},
    year = {2022},
    journal = {Physical Review E},
    author = {Philippe, Thomas and Henry, Hervé and Plapp, Mathis},
    number = {3},
    month = {9},
    volume = {106},
    publisher = {American Physical Society},
    doi = {10.1103/PhysRevE.106.034119},
    issn = {24700053},
    pmid = {36266819}
}

@article{Leo1989TheSolution,
    title = {{The effect of elastic fields on the morphological stability of a precipitate grown from solid solution}},
    year = {1989},
    journal = {Acta Metallurgica},
    author = {Leo, P H and Sekerka, R F},
    number = {12},
    pages = {3139--3149},
    volume = {37}
}

@article{Yoo1995TheSuperalloy,
    title = {{The effect of elastic misfit strain on the morphological evolution of y'-precipitates in a model Ni-base superalloy}},
    year = {1995},
    journal = {Metals And Materials},
    author = {Yoo, Y S and Yoon, D Y and Henry, M F},
    pages = {47--61},
    volume = {I, No. I}
}

@techreport{Thompson1994ThePrecipitate,
    title = {{The Equillibrium Shape of a misfitting precipitate}},
    year = {1994},
    booktitle = {{\~{}} Pergamon Acta metall, mater},
    author = {Thompson, M E and Su, C S and Voorhees, P W},
    number = {6},
    pages = {2107--2122},
    volume = {42}
}

@article{Yeom1993TheCoarsening,
    title = {{The morphological changes of y' precipitates in a Ni-8AI (Wt Pct) alloy during their coarsening}},
    year = {1993},
    journal = {Metallurgical Transactions A},
    author = {Yeom, S J and Yoon, D Y and Henry, M F},
    pages = {1975--1981},
    volume = {42A}
}

@article{Momma2008VESTA:Analysis,
    title = {{VESTA: A three-dimensional visualization system for electronic and structural analysis}},
    year = {2008},
    journal = {Journal Of Applied Crystallography},
    author = {Momma, Koichi and Izumi, Fujio},
    number = {3},
    month = {4},
    pages = {653--658},
    volume = {41}
}
\end{document}